\documentclass[final,twocolumn,3p]{elsarticle}
\usepackage{amssymb}
\usepackage{graphicx}
\usepackage{geometry}
\usepackage[dvipsnames,table]{xcolor}
\usepackage{slashed}
\usepackage{hyperref} 
\usepackage{xspace}
\usepackage[tight]{subfigure}
\usepackage{amsmath}
\usepackage{amsfonts}
\usepackage{mathrsfs}
\usepackage{comment}
\usepackage{afterpage}
\usepackage{verbatim}
\usepackage{booktabs}
\usepackage{array}
\usepackage[title,titletoc]{appendix}
\usepackage{tabularx}
\newcolumntype{C}[1]{>{\centering\arraybackslash}p{#1}}
\def\reffi#1{\mbox{Figure~\ref{#1}}}

\def\refse#1{\mbox{Section~\ref{#1}}}

\def\citere#1{\mbox{Ref.~\cite{#1}}}

\newcommand{\newc}{\newcommand}
\newc{\nnb}{\nonumber}
\newc{\beqn}{\begin{eqnarray}}
\newc{\eeqn}{\end{eqnarray}}
\newc{\beq}{\begin{equation}}
\newc{\eeq}{\end{equation}}
\newc{\bit}{\begin{itemize}}
\newc{\eit}{\end{itemize}}
\newc{\ben}{\begin{enumerate}}
\newc{\een}{\end{enumerate}}
\newc{\bce}{\begin{center}}
\newc{\ece}{\end{center}}
\newc{\bfi}{\begin{figure}}
\newc{\efi}{\end{figure}}

\newcommand{\ri}{\mathrm i}

\newcommand{\rT}{{\mathrm{T}}}

\newcommand{\rL}{{\mathrm{L}}}

\newcommand{\ie}{\emph{i.e.}\ }

\newcommand{\GeV}{\ensuremath{\,\text{GeV}}\xspace}
\newcommand{\TeV}{\ensuremath{\,\text{TeV}}\xspace}

\newcommand{\PH}{\ensuremath{\text{H}}\xspace}

\newcommand{\Pp}{\ensuremath{\text{p}}}
\newcommand{\Pe}{\ensuremath{\text{e}}\xspace}
\newcommand{\Pb}{\ensuremath{\text{b}}\xspace}

\newcommand{\Pt}{\ensuremath{\text{t}}\xspace}

\newcommand{\Pg}{\ensuremath{\text{g}}}

\newcommand{\PW}{\ensuremath{\text{W}}\xspace}
\newcommand{\PZ}{\ensuremath{\text{Z}}\xspace}

\newcommand{\Mt}{\ensuremath{m_\Pt}\xspace}
\newcommand{\MH}{\ensuremath{M_\PH}\xspace}
\newcommand{\MWOS}{\ensuremath{M_\PW^\text{OS}}\xspace}
\newcommand{\MW}{\ensuremath{M_\PW}\xspace}
\newcommand{\MZOS}{\ensuremath{M_\PZ^\text{OS}}\xspace}
\newcommand{\MZ}{\ensuremath{M_\PZ}\xspace}

\newcommand{\Gt}{\ensuremath{\Gamma_\Pt}\xspace}
\newcommand{\GH}{\ensuremath{\Gamma_\PH}\xspace}
\newcommand{\GZ}{\ensuremath{\Gamma_\PZ}\xspace}
\newcommand{\GZOS}{\ensuremath{\Gamma_\PZ^\text{OS}}\xspace}
\newcommand{\GW}{\ensuremath{\Gamma_\PW}\xspace}
\newcommand{\GWOS}{\ensuremath{\Gamma_\PW^\text{OS}}\xspace}

\newcommand{\GF}{\ensuremath{G_\mu}}

\newcommand{\pt}[1]{\ensuremath{p_{\text{T},#1}}\xspace}

\newcommand{\recola}{{\sc Recola}\xspace}

\newcommand{\mocanlo}{{\sc MoCaNLO}\xspace}
\newcommand{\bbmc}{{\sc BBMC}\xspace}
\newcommand{\collier}{{\sc Collier}\xspace}

\newcolumntype{.}{D{.}{.}{-1}}
\newcolumntype{d}[1]{D{.}{.}{#1}}
\colorlet{tableoverheadcolor}{gray!37.5}
\colorlet{tableheadcolor}{gray!25}
\colorlet{tablerowcolor}{gray!12.5}

\def\draftdate{\relax}
\def\mda{\relax}
\def\mua{\relax}
\def\mla{\relax}
\def\draft{
\def\thtystars{******************************}
\def\sixtystars{\thtystars\thtystars}
\typeout{}
\typeout{\sixtystars**}
\typeout{* Draft mode!
         For final version remove \protect\draft\space in source file *}
\typeout{\sixtystars**}
\typeout{}
\def\draftdate{\today}
\def\mua{\marginpar[\boldmath\hfil$\uparrow$]%
                   {\boldmath$\uparrow$\hfil}\color{black}%
                    \typeout{marginpar: $\uparrow$}\ignorespaces}
\def\mda{\color{red}\marginpar[\boldmath\hfil$\downarrow$]%
                   {\boldmath$\downarrow$\hfil}%
                    \typeout{marginpar: $\downarrow$}\ignorespaces}
\def\mla{\marginpar[\boldmath\hfil$\rightarrow$]%
                   {\boldmath$\leftarrow $\hfil}%
                    \typeout{marginpar: $\leftrightarrow$}\ignorespaces}
\def\Mua{\marginpar[\boldmath\hfil$\Uparrow$]%
                   {\boldmath$\Uparrow$\hfil}\color{black}%
                    \typeout{marginpar: $\uparrow$}\ignorespaces}
\def\Mda{\color{red}\marginpar[\boldmath\hfil$\Downarrow$]%
                   {\boldmath$\Downarrow$\hfil}%
                    \typeout{marginpar: $\downarrow$}\ignorespaces}
\def\Mla{\marginpar[\boldmath\hfil\textcolor{red}{$\Rightarrow$}]%
                   {\boldmath\textcolor{red}{$\Leftarrow $}\hfil}%
                    \typeout{marginpar: $\leftrightarrow$}\ignorespaces}
\overfullrule 5pt
\oddsidemargin 15mm
\marginparwidth 29mm
}

\newcommand{\ww}{\Pe^+\nu_{\Pe}\mu^-\bar{\nu}_{\mu}}
\newcommand{\brabar}{\scalebox{.3}{(\,}\raisebox{-1.9pt}{--}\scalebox{.3}{\,)}} 

\journal{Physics Letters B}

\begin{document}
\begin{frontmatter} 
\title{\hfill ~\\[-30mm]
\phantom{h} \hfill\mbox{\small COMETA-2023-01, MPP-2023-268}\\[1cm]
\vspace{13mm}   {NLO EW corrections to polarised W$^+$W$^-$ production and decay at the LHC}}
  \author[label1]{Ansgar Denner}\ead{ansgar.denner@uni-wuerzburg.de}
  \author[label1]{Christoph Haitz}\ead{christoph.haitz@uni-wuerzburg.de} 
  \author[label2]{Giovanni Pelliccioli} \ead{gpellicc@mpp.mpg.de}
  \address[label1]{University of W\"urzburg, Institut f\"ur Theoretische Physik und Astrophysik,
    Emil-Hilb-Weg 22, 97074 W\"urzburg (Germany)}
  \address[label2]{Max-Planck-Institut f\"ur Physik, Bolzmannstrasse 8, 85748 Garching (Germany)}
\begin{abstract}
  In this letter we present results for next-to-leading-order
  electroweak corrections to doubly polarised $\PW^+\PW^-$ production
  at the LHC in the fully leptonic decay channel.  We model the
  production and the decay of two polarised $\PW$~bosons in the
  double-pole approximation, including factorisable real and virtual
  electroweak corrections, and separating polarisation states at
  amplitude level.  We obtain integrated and differential predictions
  for polarised signals in a realistic fiducial setup.
\end{abstract}
\end{frontmatter}

%%%%%%%%%%%%%%%%%%%%%%%%%%%%%%%%%%%%%%%%%%%%%%%%%%%%%%%%%%%%%%%%%%%%%%%%%%%%%%%%%%%%%%%%%%%%%%%%%%
\section{Introduction}\label{sec:intro}
\noindent
Separating polarisation modes of $\PW$ and $ \PZ$ bosons
and extracting the longitudinal one represents an important step towards the
complete understanding of the interplay between the electroweak (EW)
and Higgs sector of the Standard Model (SM).
Any deviation from the SM prediction for the production rate of longitudinally
polarised EW bosons in LHC processes would signal the presence of
new-physics effects pointing to a modified structure of the EW-symmetry-breaking
mechanism compared to the SM one.

Though challenging, the extraction of polarisation fractions of EW bosons from 
LHC data has already been achieved with the Run-2 dataset
\cite{Aaboud:2019gxl,Sirunyan:2020gvn,CMS:2021icx,ATLAS:2022oge,ATLAS:2023zrv},
and more results are expected from the Run-3 and High-Luminosity stages.
The current analysis strategies for diboson inclusive production
\cite{Aaboud:2019gxl,CMS:2021icx,ATLAS:2022oge,ATLAS:2023zrv}
and scattering \cite{Sirunyan:2020gvn} are based on data fits performed
with independent templates for the various polarised signals.
This is made possible by the definition of polarised cross sections, \ie
cross sections for fixed polarisation of intermediate EW bosons, and their
numerical simulation with Monte Carlo codes.
The recent theoretical progress has lead to accurate and precise
SM predictions for inclusive polarised diboson production in the fully leptonic decay channel,
achieving next-to-leading order (NLO) QCD accuracy for all diboson channels
\cite{Denner:2020bcz,Denner:2020eck,Denner:2022riz,Hoppe:2023uux},
 next-to-next-to-leading order (NNLO) QCD accuracy for $\PW^+\PW^-$ \cite{Poncelet:2021jmj},
and NLO QCD+EW accuracy for $\PZ\PZ$ \cite{Denner:2021csi} and $\PW\PZ$
\cite{Le:2022ppa,Le:2022lrp,Dao:2023pkl} production.

The inclusive production of a pair of leptonically  decaying $\PW$
bosons at the  LHC is the most challenging
diboson channel from the experimental point of view, owing to the large
top-quark backgrounds and the cumbersome
reconstruction of the final state with two undetected neutrinos.
The SM differential cross section for off-shell $\PW^+\PW^-$
production is known up to NNLO QCD
\cite{Caola:2015rqy,Grazzini:2016ctr,Kallweit:2019zez}
and NLO EW \cite{Billoni:2013aba,Biedermann:2016guo,Kallweit:2019zez}
perturbative accuracy, also matched to parton shower
\cite{Chiesa:2020ttl,Brauer:2020kfv,Lombardi:2021rvg}.
In the case of intermediate polarised bosons, the NLO \cite{Denner:2020bcz}
and NNLO \cite{Poncelet:2021jmj} QCD corrections are known, and the matching
of NLO QCD ones to a parton shower
\looseness -1
has been achieved recently \cite{Pelliccioli:2023zpd}.

In this work we extend the SM modelling of doubly polarised
$\PW^+\PW^-$ production to the inclusion of NLO EW corrections in the
decay channel with two charged leptons of different flavours.

This letter is organised as follows. In \refse{sec:calc} we describe the main
features of the NLO EW calculation. The setup for numerical simulations is summarised
in \refse{sec:setup}, and the corresponding results are discussed in \refse{sec:res}.
In \refse{sec:concl} we draw our conclusions.
%%%%%%%%%%%%%%%%%%%%%%%%%%%%%%%%%%%%%%%%%%%%%%%%%%%%%%%%%%%%%%%%%%%%%%%%%%%%%%%%%%%%%

\section{Details of the calculation}\label{sec:calc}
\noindent
We consider the production and decay of two $\PW$~bosons at NLO EW
accuracy, specifically
\beq
\Pp\Pp\rightarrow \ww+ X\,.
\eeq
In the five-flavour scheme, the channels
\beqn
q\bar{q},\, \Pb\bar{\Pb},\, \gamma\gamma&\rightarrow& \ww\,,
\eeqn
 ($q={\rm u,d,s,c}$) are present at LO, and at NLO EW accuracy the contributing real channels read
\beqn
  q\bar{q},\, \Pb\bar{\Pb}\,, \gamma\gamma &\rightarrow& \ww\gamma\,,\nnb\\
  \gamma \overset{\brabar}{q}&\rightarrow &\ww\,\overset{\brabar}{q}\,,\nnb\\
  \gamma\overset{\brabar}{\Pb}&\rightarrow &\ww\overset{\brabar}{\Pb}\,.
\eeqn
In the full off-shell calculation, the complete set of real and virtual diagrams is taken into account.

For unpolarised and doubly polarised signals, we carry out the calculation in the double-pole approximation (DPA)
%\cite{Stuart:1991cc,Stuart:1991xk,Aeppli:1993cb,Aeppli:1993rs,Denner:2000bj},
\cite{Stuart:1991xk,Aeppli:1993cb,Aeppli:1993rs,Denner:2000bj},
including only factorisable real and virtual EW corrections.
The DPA approach has already been used for the simulation of polarised intermediate EW bosons
in diboson LHC processes at NLO QCD \cite{Denner:2020bcz,Denner:2020eck,Denner:2022riz},
NNLO QCD \cite{Poncelet:2021jmj}, and NLO EW \cite{Denner:2021csi,Le:2022ppa,Le:2022lrp} accuracy.

The $\gamma\Pb\,(\gamma\bar{\Pb})$ channels embed (anti)top quarks
in the $s$~channel, and should therefore be regarded as an irreducible
background to the $\PW^+\PW^-$ EW production. 
Therefore, while included as a contribution to the full off-shell
calculation as a reference, these contributions are excluded from the DPA calculations
(polarised and unpolarised), \ie we assume a perfect b-jet veto.

In the DPA, the Born-like contributions (Born, virtual, and subtraction counterterms)
are characterised by two $\PW$~bosons undergoing two-body decays,
\beq
q\bar{q},\,\Pb\bar{\Pb},\,\gamma\gamma\rightarrow\PW^+(\Pe^+\nu_{\Pe})\,\PW^-(\mu^-\bar{\nu}_\mu),
\eeq
where the notation $\PW^+(\Pe^+\nu_{\Pe})$ denotes a
$\PW^+$~boson decaying into $\Pe^+\nu_{\Pe}$.
The same holds for real corrections with an additional particle, a photon or a quark, radiated off
the production part of the process,
\beqn
q\bar{q},\Pb\bar{\Pb},\gamma\gamma&\rightarrow&\PW^+(\Pe^+\nu_{\Pe})\,\PW^-(\mu^-\bar{\nu}_\mu)\,\gamma\,,\nnb\\
\gamma \overset{\brabar}{q}
&\rightarrow& \PW^+(\Pe^+\nu_{\Pe})\,\PW^-(\mu^-\bar{\nu}_\mu)\,\overset{\brabar}{q}\,,\nnb\\
\gamma \overset{\brabar}{b}
&\rightarrow& \PW^+(\Pe^+\nu_{\Pe})\,\PW^-(\mu^-\bar{\nu}_\mu)\,\overset{\brabar}{b}\,.
\eeqn
Following \citere{Denner:2021csi}, these contributions are treated with the DPA(2,2) mapping.
A photon can be also emitted off the decay products of a $\PW$~boson, leading to a resonant structure with
one two-body and one three-body decay,
\beqn
q\bar{q},\Pb\bar{\Pb},\gamma\gamma&\rightarrow&\PW^+(\Pe^+\nu_{\Pe}\gamma)\,\PW^-(\mu^-\bar{\nu}_\mu)\,,\nnb\\
q\bar{q},\Pb\bar{\Pb},\gamma\gamma&\rightarrow&\PW^+(\Pe^+\nu_{\Pe})\,\PW^-(\mu^-\bar{\nu}_\mu\gamma)\,.
\eeqn
According to the notation of \citere{Denner:2021csi}, these contributions are treated with the DPA(3,2) and DPA(2,3) mappings, respectively.

The infrared singularities of QED origin are subtracted in the dipole formalism \cite{Catani:1996vz,Dittmaier:1999mb,Catani:2002hc,Dittmaier:2008md,Schonherr:2017qcj}.
Compared to the full off-shell calculation, where emitters and spectators of the dipoles are charged massless particles in the initial or final state,
the DPA calculations require a tailored dipole selection, owing to the separate treatment of the production and decay of the resonances.
%At variance with the $\PZ$-boson case \cite{Denner:2021csi}, 
This procedure is especially delicate for the soft-photon singularities
that arise from the emission of photons off $\PW$~bosons whose momenta have been projected on mass shell through a DPA mapping.
In order to further detail the DPA approach used for our calculation,
we consider the real partonic process $q\bar{q}\rightarrow\ww\gamma$. 
The contributing diagrams in the DPA can feature a photon that is radiated off the production part of the amplitude, from a $\PW$-boson propagator,
or from the decay part of the amplitude. 
\begin{figure*}[h!]
   \centering
   \begin{tabular}{ccc}
     \subfigure[\label{fig:diagProd}]{\includegraphics[scale=0.38]{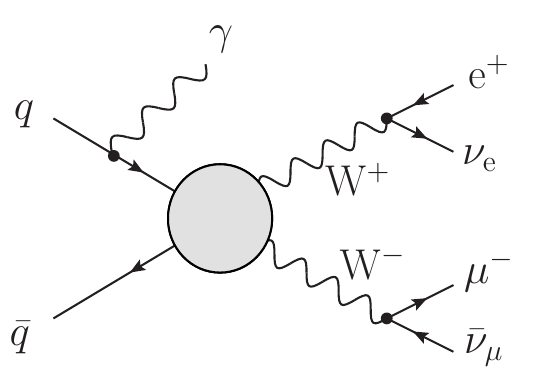}} &
     \subfigure[\label{fig:diagProp}]{\includegraphics[scale=0.38]{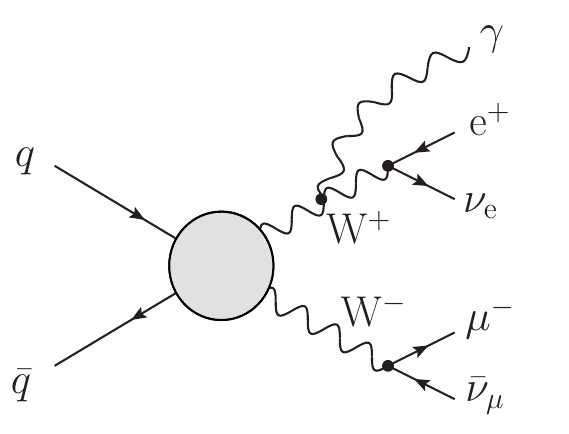}} &
     \subfigure[\label{fig:diagDec}]{\includegraphics[scale=0.38]{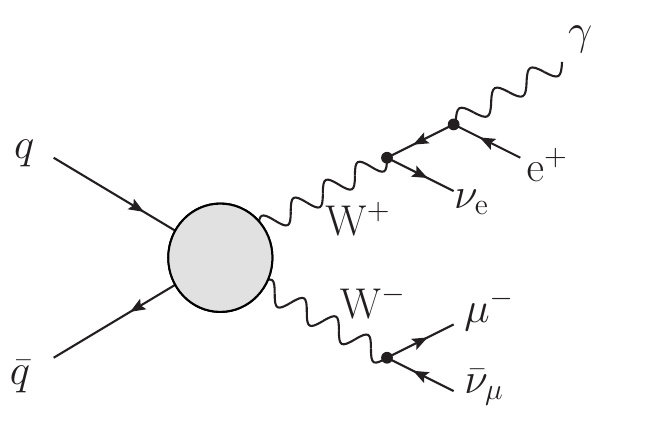}} \\
   \end{tabular}
   \caption{Sample photon-radiation diagrams contributing to
     $\PW^+\PW^-$ production and decay at NLO EW.}\label{fig:diags} 
\end{figure*}
Since diagrams like the one depicted in \reffi{fig:diagProd} contain a
photon radiated off initial-state (IS) quarks or antiquarks, they 
only contribute to the production process of two $\PW$~bosons. These
diagrams give rise to soft and collinear singularities, which in the dipole formalism 
are absorbed by IS--IS and IS--FS dipoles, where either another quark
or antiquark (IS--IS) or one of the two $\PW$~bosons (IS--FS) plays the role of the spectator.
The second class of diagrams [\reffi{fig:diagProp}] is characterised by a photon emitted off an  $s$-channel $\PW$~boson
and embeds singularities associated both with the production side and with the decay side of the amplitude, treated separately
my means of a partial fractioning
\cite{Dittmaier:2015bfe,Denner:2019vbn}. This gives rise to FS--IS
($\PW$~boson as emitter, incoming 
parton as spectator) and FS--FS dipoles (one $\PW$~boson is the emitter, the other $\PW$ is the
spectator) to absorb production-level singularities, and a decay dipole for the decay-level singularities.
The last class of diagrams [\reffi{fig:diagDec}] contains a photon
emitted from a charged lepton, whose singularity is absorbed by a decay dipole.
Following the strategy proposed in \citere{Le:2022ppa}, for the production-level dipoles with a $\PW$~boson as emitter and/or spectator (FS--FS, FS--IS, IS--FS) the massive-fermion dipoles
\cite{Catani:2002hc} are employed. This is possible, since owing to the
finite mass of the $\PW$~boson no spin-dependent collinear
singularities are present in the photon radiation from the $\PW$~bosons.
Compared to $\PW\PZ$ inclusive production \cite{Le:2022lrp,Le:2022ppa}, the novel structures needed for this calculation are the
FS--FS dipoles with equal masses ($\MW$) that we implemented as a simplified version of the corresponding structures in \citere{Catani:2002hc}.
For the $\PW$-boson decay, we have devised a single dipole that reproduces the exact structure of the real $\PW\rightarrow\ell\nu_\ell\gamma$ matrix element
and that could be analytically integrated in $4-2\epsilon$ dimensions. This dipole absorbs the singularities related to photon emission
both from the $\PW$ boson and from the decay lepton.

The sum of integrated counterterms for the production part (taken from \citere{Catani:2002hc}) and decay (integrated version of our $\PW$-decay kernel)
has been proven to reproduce the explicit infrared poles of the factorisable virtual corrections. The contribution of
non-factorisable soft-photon corrections of virtual origin is expected
to be small and to cancel to a large extent against the corresponding real
corrections \cite{Denner:1997ia}. Therefore we exclude them from this
work and leave them for future investigation.

The calculation strategy described above for real and virtual contributions 
as well as for the local and integrated counterterms
has been applied to the unpolarised process
and to the doubly polarised ones. This allows the selection of
individual polarisation contributions in the two $\PW$-propagator numerators, following
the general strategy proposed in \citere{Denner:2020bcz} and already applied at NLO EW to  
$\PZ\PZ$ \cite{Denner:2021csi} and $\PW\PZ$ inclusive production \cite{Le:2022ppa,Le:2022lrp,Dao:2023pkl}.

At variance with the choice made in other fixed-order results in the literature \cite{Denner:2020bcz,Poncelet:2021jmj}, the polarisation states
for the two $\PW$~bosons are defined in the diboson centre-of-mass
(CM) frame, which is regarded as the most natural Lorentz frame for the
definition of vector-boson polarisations in diboson
processes \cite{Denner:2020eck,Denner:2021csi,Le:2022ppa,Le:2022lrp,Dao:2023pkl}.

The calculation has been performed independently with the
\bbmc and \mocanlo Monte Carlo codes, both of which have already been
used for the simulation of intermediate polarised bosons at NLO accuracy
\cite{Denner:2020bcz,Denner:2020eck,Denner:2021csi,Denner:2022riz}.
The two codes have been interfaced with the latest release (1.4.4) of the \recola library
\cite{Actis:2012qn,Actis:2016mpe} that enables the calculation of tree-level and
one-loop amplitudes with fixed polarisation states for intermediate resonances.
The reduction and integration of one-loop amplitudes in \recola is achieved through the \collier
library \cite{Denner:2016kdg}. A number of checks have been performed to verify the correctness of fixed-helicity
amplitudes by means of variations of the UV regulator in \recola, comparisons against analytic results, and independent
numerical results obtained with {\sc MadLoop} \cite{Hirschi:2011pa}.
All integrated and differential results provided in this letter are
computed with \mocanlo and have been checked against \bbmc,
finding agreement within numerical-integration uncertainties.

\section{Setup}\label{sec:setup}
\noindent
The simulations are performed at a centre-of-mass energy of $\sqrt{s} = 13.6\TeV$ for proton--proton collisions at the LHC.
The on-shell masses and widths of the EW bosons have been set to the values \cite{Workman:2022ynf},
\begin{align}
 \MZOS  &{}=  91.1876\GeV,\quad &
 \GZOS  &{}=  2.4952\GeV,\nonumber \\
 \MWOS  &{}=  80.377\GeV,\quad &
 \GWOS  &{}=  2.085\GeV\,,
\end{align}
and then converted to their corresponding pole values \cite{Bardin:1988xt}.
The top-quark and Higgs-boson mass and width are fixed as \cite{Workman:2022ynf},
\begin{align}
\Mt &{}= 172.69\GeV,\quad &
\Gt &{}= 1.42\GeV, \nnb\\ 
\MH &{}= 125.25\GeV,\quad & %125.11 in Workman:2022ynf \cite{ATLAS:2023dnm,ATLAS:2023oaq}
\GH &{}= 0.0041\GeV\,. % 0.0045 in Workman:2022ynf \cite{ATLAS:2023dnm,ATLAS:2023oaq}
\end{align}
%% \begin{align}
%% \qquad  \MZOS &{}=  91.1876\GeV, &
%%   \GZOS &{}=  2.4952\GeV,  \qquad \nonumber \\
%%   \MWOS &{}=  80.377\, \GeV, &
%%   \GWOS &{}=  2.085\,\GeV\,,
%% \end{align}
%% and then converted to their corresponding pole values \cite{Bardin:1988xt}.
%% The top-quark and Higgs-boson mass and width are fixed as \cite{Workman:2022ynf},
%% \begin{align}
%%   \Mt &{}= 172.69\GeV, &
%%   \Gt &{}= \,  1.42\GeV, \nnb\\
%%   \MH &{}=   125.25 \, \GeV, &  %125.11 in Workman:2022ynf \cite{ATLAS:2023dnm,ATLAS:2023oaq}
%%   \GH &{}=  \, 0.0041\,\GeV\,. % 0.0045 in Workman:2022ynf \cite{ATLAS:2023dnm,ATLAS:2023oaq}
%% \end{align}
The $G_\mu$ scheme \cite{Denner:2000bj,Dittmaier:2001ay} is employed to determine the EW coupling.
In formulas, 
\begin{align}\label{eq:alphadef}
  &\alpha =
  \frac{\sqrt{2}}{\pi}\,G_\mu\,\left|\hat{M}_{\PW}^2\,
%  \frac{\sqrt{2}}{\pi}\,G_\mu\,\left|(\MW^2-\ri \MW\GW)\,
  \left(
  1 - \frac{\hat{M}_{\PW}^2}{\hat{M}_{\PZ}^2}
%  1 - \frac{\MW^2-\ri \MW\GW}{\MZ^2-\ri \MZ\GZ}
  \right)\right|
\end{align}
for the full off-shell calculation in the complex-mass scheme \cite{Denner:2005fg,Denner:2006ic,Denner:2019vbn},
where $\hat{M}_V^2=M_V^2-\ri M_V\Gamma_V$ ($V=\PW,\,\PZ$), while
\begin{align}
  &\alpha =
  \frac{\sqrt{2}}{\pi}\,G_\mu\,\MW^2\,
  \left(
  1 - \frac{\MW^2}{\MZ^2}
  \right)
\end{align}
for the DPA calculations. The Fermi constant is set to
\begin{equation}
  \GF = 1.16638\times 10^{-5}\GeV^{-2}\,,
\end{equation}
and $\MZ,\MW$ and $\GW, \GZ$ represent the pole values for the weak-boson masses and widths, respectively.

We perform the calculation in the five-flavour scheme, including partonic channels induced by bottom (anti)quarks and photons.
The PDF set \sloppy\texttt{NNPDF31\_nnlo\_as\_0118\_luxqed}~\cite{NNPDF:2017mvq,Bertone:2017bme}
has been utilised through the \textsc{Lhapdf} interface \cite{Buckley:2014ana}.
The renormalisation and factorisation scales are both set to the $\PW$~pole mass,
\begin{equation}
\label{eq:scale}
\mu_{\rm R} =
\mu_{\rm F} =
\MW \,.
\end{equation}

\noindent
The selections used throughout this paper mimic those of a recent CMS measurement \cite{CMS:2020mxy}
(dubbed \emph{sequential-cut selections} therein).
The charged leptons are dressed with photon radiation according to
anti-$k_{\rm T}$ clustering algorithm \cite{Cacciari:2008gp} with
resolution radius $R=0.1$. %\GP{Here CMS would have cone dressing with 0.3 and 0.4 radius for the electron and muon, respectively.}
The final state must satisfy
\beqn\label{eq:fiducial}
&&\pt{\ell_1}>25\GeV\,,\quad
\pt{\ell_2}>20\GeV\,,\nnb\\
&&|\eta_{\Pe^+}|<2.5\,,\quad
|\eta_{\mu^-}|<2.4\,,\nnb\\
&&\pt{\Pe^+\mu^-}>30\GeV\,,\quad
M_{\Pe^+\mu^-}>20\GeV\,,\nnb\\
&&\pt{\rm miss}>20\GeV\,,
\eeqn
where $\ell_{1,2}$ are the leading and subleading charged leptons
ordered according to their transverse momenta.

Throughout the whole paper we use the labels $\rL$ and $\rT$ for longitudinal
and transverse polarisation, respectively. When discussing doubly
polarised states (LL, LT, TL, and TT), the first index refers to the $\PW^+$
boson, the second one to the $\PW^-$~boson. 

\section{Results}\label{sec:res}
In this section we present integrated and differential results at NLO
EW order in the fiducial setup defined in Eq.~\eqref{eq:fiducial}.
Since we are interested in assessing the impact of EW radiative corrections, we do not show QCD-scale uncertainties which in our
calculation would only come from factorisation-scale variations. A realistic estimate of QCD-scale uncertainties would require at least the
inclusion of NLO QCD corrections, which has been recently carried out \cite{Pelliccioli:2023zpd} in a similar setup as the one considered in this work,
finding 3--5\% uncertainties with mild differences among various polarised states.

In Table~\ref{tab:intEW} we show fiducial cross sections at LO and NLO
EW accuracy for the full, unpolarised, and doubly polarised $\PW^+\PW^-$ signals. 
The results are presented both with and without the $\Pb\bar{\Pb}$ channel. 
In the case of the full calculation, we also assess the impact
of the $\gamma\Pb\,(\gamma\bar{\Pb})$ channels, which differ from the corresponding ones with light quarks by the presence of (anti)top-quark
propagators in the $s$~channel.
\begin{table*}[t]
     \begin{center}
       \begin{tabular}{ccccc}%
         \hline\rule{0ex}{2.7ex}
         \cellcolor{blue!14} state  & \cellcolor{blue!14} $\sigma_{\rm LO}$ [fb] & \cellcolor{blue!14} $\sigma_{\rm NLO\,EW}$ [fb] & \cellcolor{blue!14} $\delta_{\rm EW}[\%]$ & \cellcolor{blue!14} $f_{\rm NLO\,EW}[\%]$\\
\hline\\[-0.3cm]
\multicolumn{5}{c}{\cellcolor{green!9} $\Pb\bar{\Pb},\, \gamma\Pb,\,\gamma\bar{\Pb}$ excluded}\\
\hline\\[-0.3cm]
full &   $  254.79(2)  $  &      $ 249.88(9)   $ &  $     -1.93  $  & $103.5$ \\[0.1cm]
unp. &   $  245.79(2)  $  &      $ 241.48(2)   $ &  $     -1.75  $  & $100$ \\[0.1cm]
LL   &   $  18.752(2)  $  &      $ 18.510(2)   $ &  $     -1.30  $  & $7.7$ \\[0.1cm]
LT   &   $  32.084(3)  $  &      $ 32.043(3)   $ &  $     -0.13  $  & $13.3$ \\[0.1cm]
TL   &   $  33.244(5)  $  &      $ 33.155(5)   $ &  $     -0.27  $  & $13.7$ \\[0.1cm]
TT   &   $  182.17(2)  $  &      $ 177.83(2)   $ &  $     -2.38  $  & $73.6$ \\[0.1cm]
int.  &  $  -20.46(3)  $  &      $-20.1(1)     $ &  $     -1.96  $  & $-8.3$ \\[0.1cm] 
\hline\\[-0.3cm]
\multicolumn{5}{c}{\cellcolor{green!9} $\Pb\bar{\Pb}$ included, $\gamma\Pb,\,\gamma\bar{\Pb}$ excluded}\\
\hline\\[-0.3cm]
full  &  $  259.02(2) $   &      $ 253.95(9)  $  &  $     -1.96  $ &  $103.4$ \\[0.1cm]
unp.  &  $  249.97(2) $   &      $ 245.49(2)  $  &  $     -1.79  $ &  $100.0$ \\[0.1cm]
LL    &  $  21.007(2) $   &      $ 20.663(2)  $  &  $     -1.64  $ &  $8.4$ \\[0.1cm]
LT    &  $  33.190(3) $   &      $ 33.115(3)  $  &  $     -0.23  $ &  $13.5$ \\[0.1cm]
TL    &  $  34.352(5) $   &      $ 34.230(5)  $  &  $     -0.35  $ &  $13.9$ \\[0.1cm]
TT    &  $  182.56(2) $   &      $ 178.21(3)  $  &  $     -2.38  $ &  $72.6$ \\[0.1cm]
int.  &  $  -21.14(5) $   &      $  -20.6(2)  $  &  $     -2.45  $ &  $-8.4$ \\[0.1cm]
\hline\\[-0.3cm]
\multicolumn{5}{c}{\cellcolor{green!9} $\Pb\bar{\Pb},\, \gamma\Pb,\,\gamma\bar{\Pb}$ included}\\
\hline\\[-0.3cm]
full  &  $  259.02(2) $   &      $ 265.59(9) $   &  $     +2.54  $ & -\\[0.1cm]
\hline
       \end{tabular}
     \end{center}
     \caption{Fiducial cross sections (in fb) at LO and NLO EW for full, unpolarised, and doubly polarised $\PW^+\PW^-$ production at the LHC in
       the fully leptonic decay channel. Absolute numbers in
       parentheses are numerical integration uncertainties. The value $\delta_{\rm EW}$ (in percentage)
       is computed as the EW correction relative to the LO result. The values $f_{\rm NLO\,EW}$ are fractions of NLO EW cross sections over
       the NLO EW unpolarised result. The $\gamma\Pb,\,\gamma\bar{\Pb}$ contributions are only included in the full calculation (last row).
       The interference (int.) is evaluated as the difference between the unpolarised and the sum of polarised results.
     }
     \label{tab:intEW}
\end{table*}
Excluding all bottom-induced contributions, the TT state gives by far the largest fraction to the unpolarised signal, while the
purely longitudinal state amounts to $7.7\%$ of the total at NLO EW.
The off-shell effects, evaluated from the difference between the full and unpolarised cross sections, are
at the $3.5\%$ level, in agreement with the intrinsic uncertainty of the DPA. The striking effect is the
size of interferences among polarisation states ($-8\%$), evaluated from the difference between the unpolarised
and sum of polarised cross sections. This effect, already observed in previous $\PW^+\PW^-$ calculations
\cite{Denner:2020bcz,Poncelet:2021jmj,Pelliccioli:2023zpd}, comes from
the interplay between the left-chiral 
$\PW$-boson coupling to fermions and the application of transverse-momentum cuts on the two
charged leptons.
The NLO EW corrections are negative and different for the various polarised and unpolarised
states. In particular, their size is maximal for the TT state ($-2.4\%$), smaller for the LL one ($-1.3\%$).
Almost negligible EW corrections are found for the mixed polarisation states (LT, TL).

The inclusion of $\Pb\bar{\Pb}$ channels, which are PDF suppressed, 
% compared with light quarks,
though not changing the size of the NLO EW corrections, gives a $12\%$ increase to the LL cross section.
The corresponding contribution for mixed states is at the $3\%$ level, while it is completely negligible for the TT state.
This effect comes from the presence of a $t$-channel top quark that
leads to a different helicity structure contributing to the
LO amplitude, favouring a longitudinal mode for the $\PW$~bosons.
Including the $\Pb\bar{\Pb}$ contributions does not change the relative size of off-shell and interference effects.

For completeness, we have evaluated the full off-shell process including also $\gamma\Pb\, (\gamma\bar{\Pb})$ channels,
which are characterised by (anti)top quarks in the $s$~channel. These
photon-induced corrections account for almost $5\%$ of the
full off-shell NLO EW cross section in the five-flavour scheme. 
%Owing to their different helicity structure, 
Dominated by
the presence of (anti)top quarks, these contributions constitute an irreducible background to EW production of $\PW^+\PW^-$.
A similar reasoning holds for $\Pg\Pb,\,\Pg\bar{\Pb}$ and $\Pg\Pg$ channels that arise at NLO and NNLO in QCD, respectively.
The combination of EW and QCD corrections, although indispensable for a realistic and precise SM modelling, falls outside the
scope of this letter and is left for future work.

In order to experimentally separate polarisation states for intermediate EW bosons,
it is necessary to model precisely the polarised signals in terms of differential
observables, identifying those that provide the highest discrimination power.
We present differential results for three kinematic variables: the first one relies on
the use of neutrino momenta from Monte Carlo truth, while the second and the third ones 
are LHC observables. While we incorporate the $\Pb\bar{\Pb}$ channels in
all distributions, we do not include the
$\gamma\Pb,\,\gamma\bar{\Pb}$ channels therein, but instead show their
contribution to the full off-shell results separately. 

\begin{figure*}[t]
   \centering
   \includegraphics[scale=0.38]{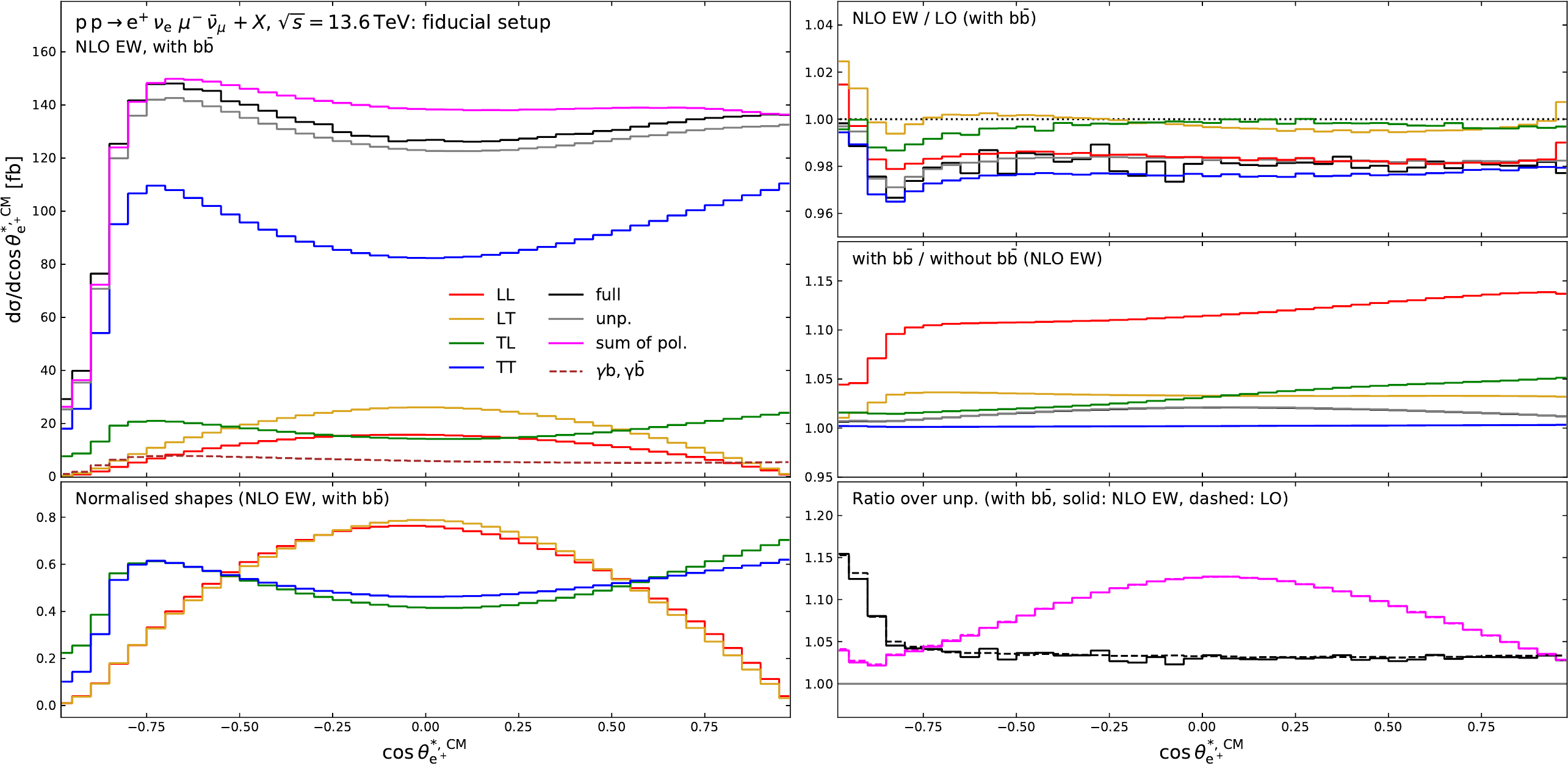}
   \caption{Distributions in the polar decay angle of the positron in the $\PW^+$~rest frame
     for $\PW^+\PW^-$ production and decay at the LHC with NLO EW accuracy.
     The setup detailed in \refse{sec:setup} is understood.
     Polarisations are defined in the diboson CM reference frame.
     Colour key: full off-shell (black), unpolarised (gray), LL (red), LT (yellow), TL (green),
     TT (blue), sum of doubly polarised (magenta), $\gamma\Pb\,(\gamma\bar{\Pb})$ contributions
     (brown, dashed).
     Left panel: absolute differential distributions (top) and normalised shapes (bottom)
     at NLO EW, including $\Pb\bar{\Pb}$ contributions. Right panel: ratios of NLO EW results
     over the LO ones, including $\Pb\bar{\Pb}$ contributions (top), ratios of NLO EW cross sections
     with and without $\Pb\bar{\Pb}$ contributions included (middle), ratios of cross sections
     over the unpolarised ones at LO and NLO EW (bottom). \label{fig:cosdec}}
\end{figure*}

In \reffi{fig:cosdec} we present distributions in the cosine of the
polar decay angle of the positron in the $\PW^+$~rest frame, which
relies on the (unphysical) reconstruction of individual $\PW$~bosons. 
As proven by the normalised shapes, this quantity gives the highest sensitivity to the polarisation
state of the $\PW^+$~boson, while it is rather agnostic of the polarisation state of the $\PW^-$
boson. The NLO EW corrections for the various states reflect those found at the integrated level
(see Table~\ref{tab:intEW}), with some deviation from the flat behaviour just in the anticollinear regime,
which is the least populated one. In this regime, the TT and TL signals would have a maximum in the
absence of cuts (driven by the favoured left-handed polarisation), while the transverse-momentum selection
on the positron distorts dramatically the shape emptying this region.
The impact of $\Pb\bar{\Pb}$ channels, negligible for the TT state, increases towards the collinear
regime for the longitudinal $\PW^-$~states (LL, TL).
The off-shell effects are flat in most of the angular range, while they increase up to $15\%$ in the
anticollinear regime. 
%This effect is driven by configurations in which the positron has moderate transverse
%momentum, and the single-resonant diagrams entering the full off-shell calculation
%give a sizeable contribution.
The interferences, already found to be large for polarisations defined in the laboratory frame
\cite{Denner:2020bcz,Poncelet:2021jmj}, are enhanced in the central region of the spectrum,
\ie $\theta_{\Pe^+}^{*,\rm CM}\approx \pi/2$, where they amount to $13\%$. These large effects are
known to come from the application of $p_{\rm T}$ cuts on the kinematics of the positron,
which prevents the interference between longitudinal and transverse modes of the $\PW^+$ boson
from integrating to zero \cite{Ballestrero:2017bxn}.

\begin{figure*}[t]
   \centering
   \includegraphics[scale=0.38]{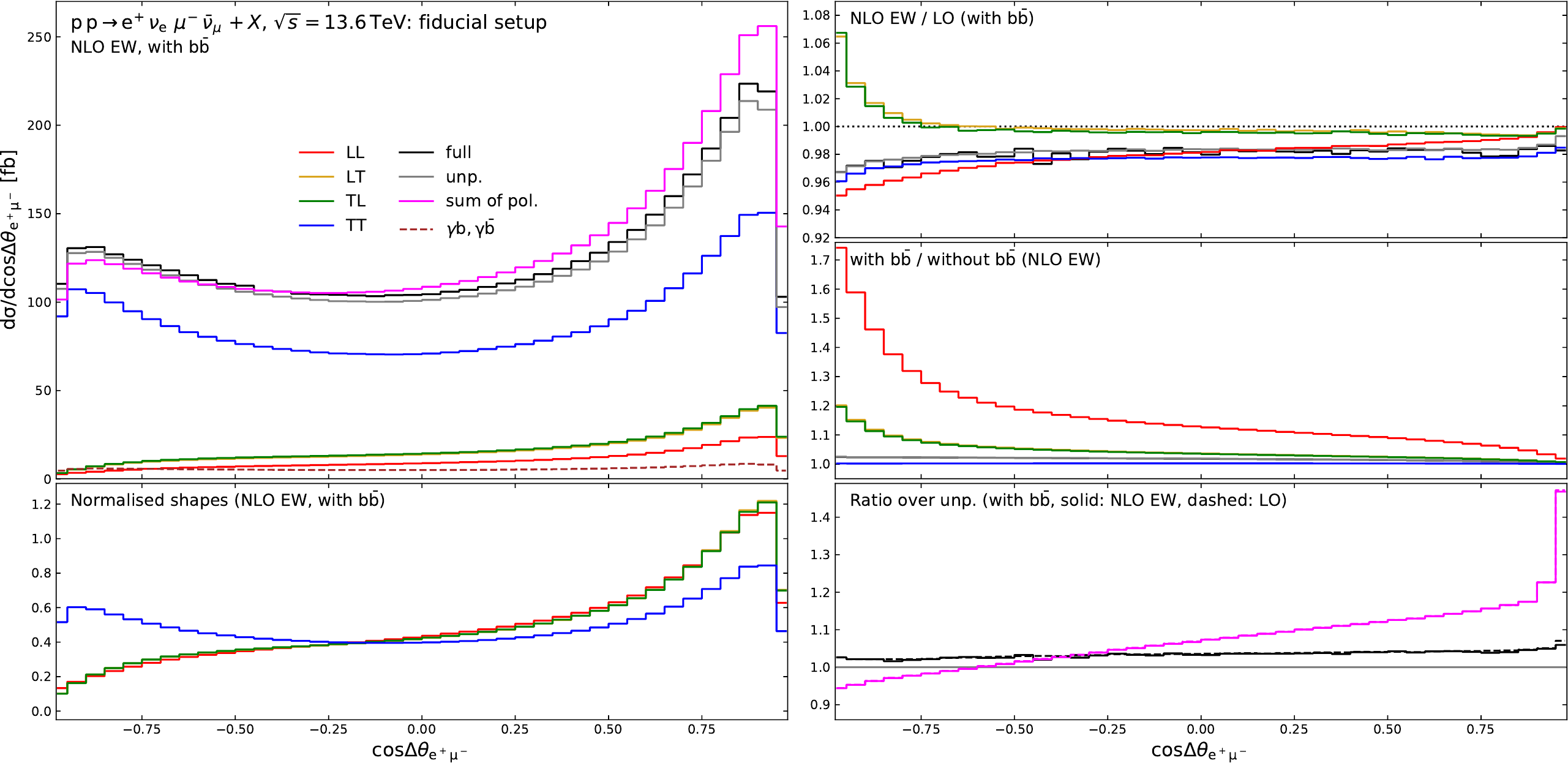}
   \caption{Distributions in the cosine of the angular separation between the positron and the muon for $\PW^+\PW^-$ production and decay at the LHC with NLO EW accuracy.
   Same structure as \reffi{fig:cosdec}.\label{fig:costh}}
\end{figure*}

In \reffi{fig:costh} we present the distributions in an angular observable that can be fully accessed at the LHC,
namely the angular separation between the two charged leptons, computed in the laboratory frame.
The shapes for the LL and mixed states clearly suggest that the two leptons are preferably
produced collinear to each other, with a sharp drop in the rightmost bin that is motivated by
the fiducial invariant-mass selection ($M_{\Pe^+\mu^-}>20\GeV$), while the anticollinear regime is
disfavoured. A different behaviour is found for the TT distribution, whose shape presents two
local maxima close to both the collinear (absolute maximum) and to the anticollinear regimes.
The NLO EW corrections for the LL state increase from $-5\%$ to $-0.1\%$ towards the collinear regime,
those for the TT state are comparably flat.  The mixed states are both characterised by positive
corrections (up to $6.5\%$) in the anticollinear region, but almost
negligible ones in the rest of the angular range.
The relative impact of $\Pb\bar{\Pb}$ contributions to the LL state is maximal in the anticollinear
region (more than $75\%$ for $\Delta\theta_{\Pe^+\mu^-}\approx \pi$)
and decreases monotonically towards the collinear region
(most populated one). A similar trend, though overall less sizeable, is found for the LT and TL states.
While the off-shell effects are flat, the interference pattern is strikingly interesting,
with a linear decrease from $+5\%$ to $-17\%$ towards the collinear region, and a deviation from
this constant slope just in the very last bins (up to $45\%$ effect).
A similar interference pattern has been observed also in the distributions in the azimuthal-angle
separation between the two charged leptons.

\begin{figure*}[t]
   \centering
   \includegraphics[scale=0.38]{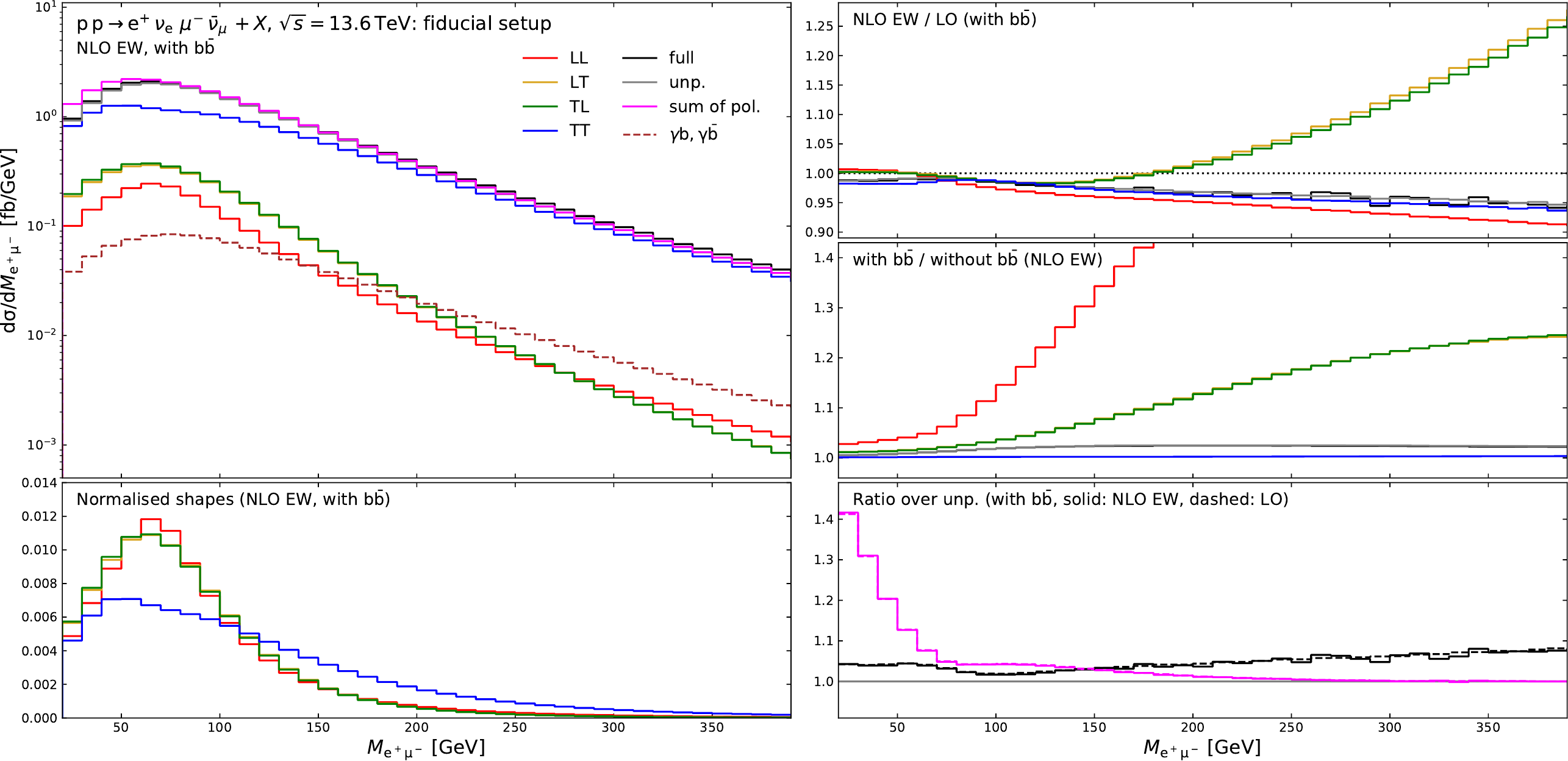}
   \caption{Distributions in the invariant mass of the positron--muon pair for $\PW^+\PW^-$ production and decay at the LHC with NLO EW accuracy.
    Same structure as \reffi{fig:cosdec}.\label{fig:mll}}
\end{figure*}

In \reffi{fig:mll} we analyse the distributions in the invariant mass of the positron--muon system.
Similarly to the discussion of the previous observable, the normalised TT shape markedly deviates from
the one found for other doubly polarised states, with a different maximum position ($M_{\Pe^+\mu^-}\approx 50\GeV$ for TT, $\approx 65\GeV$ for
other states). The TT state dominates the unpolarised cross section with a decrease by 1.5 orders of magnitude between its maximum position
and $400\GeV$. The suppression of the other states towards large
invariant-mass values is more marked, with a decrease by more than
two orders of magnitude in the same mass range.
The LL distribution exceeds the mixed ones already at a smaller
invariant mass ($\approx 270\GeV$)
than for polarised states defined in the laboratory frame \cite{Denner:2020bcz}.
The NLO EW corrections are increasingly negative for the LL and TT states towards large invariant mass, showing
the expected enhancement from large logarithms of EW origin. On the contrary, the positively increasing
EW corrections at moderate-to-large mass found for the LT and TL distributions mostly come from a LO suppression of these states.
The impact of the $\Pb\bar{\Pb}$ channel becomes very large at
moderate values of $M_{\Pe^+\mu^-}$ ($+50\%$ at $200\GeV$ for LL). 
The off-shell effects mildly vary between $3\%$ and $8\%$ in the considered range.
The interferences vanish for larger mass, while they rapidly increase
approaching the minimum mass value allowed by the selections ($40\%$
at 20\GeV cut).

\section{Conclusions}\label{sec:concl}
We have presented the first calculation of NLO EW corrections to doubly polarised $\PW^+\PW^-$ production at the LHC in the
decay channel with two opposite-sign, different-flavour leptons.
Using the double-pole approximation to extract the $\PW\PW$-resonant
contributions out of 
%the complete set of contributions to
the full off-shell cross section and selecting polarisation states for intermediate bosons in the tree-level and one-loop
amplitudes, we have calculated integrated and differential cross sections for various polarised states and drawn phenomenological
consequences relevant for the LHC Run-3 analysis programme.
We have carried out the calculation in the five-flavour scheme, finding non-negligible contributions from the
$\Pb\bar{\Pb}$-induced partonic channel, especially for the purely longitudinal state.
The NLO EW corrections at integrated level for the doubly polarised states range
from $-0.3\%$ for mixed states to $-1.7\%$ for the purely longitudinal one to $-2.4\%$ for the purely transverse one.
Much larger EW corrections characterise the tails of invariant-mass distributions, with typical
enhancement from EW logarithms in the same-polarisation states and positively increasing corrections
for the mixed states owing to a LO suppression.
Some observables, both angular and energy-dependent, have been found to have a marked discrimination power
among polarisations, and therefore to be suitable for polarised-template fits of LHC data, even in a
challenging process like $\PW^+\PW^-$.

\section*{Acknowledgements}
\noindent
The authors are grateful to Jean-Nicolas Lang and Sandro Uccirati for maintaining \recola.
GP thanks Giulia Zanderighi and Rene Poncelet for useful discussions.
The authors acknowledge support from the COMETA COST Action CA22130.
This work is supported by the German Federal Ministry for Education and Research
(BMBF) under contract no.~05H21WWCAA and by the German Research Foundation
(DFG) under reference number DFG 623/8-1.

\bibliographystyle{elsarticle-num} 
\biboptions{numbers,sort&compress}
\bibliography{polarisation}
\end{document}